\documentclass[prl,twocolumn]{revtex4}

\usepackage{graphicx}
\usepackage{amssymb,amsmath}
\begin{document}
\title{Exact solution of the one-dimensional deterministic Fixed-Energy Sandpile}
\author{Luca Dall'Asta}
\affiliation{Laboratoire de Physique Th\'eorique (UMR du CNRS 8627) - B\^atiment 210, Universit\'e de Paris-Sud, 91405 ORSAY Cedex (France)} 

\begin{abstract}
In reason of the strongly non-ergodic dynamical behavior, universality properties of deterministic Fixed-Energy Sandpiles are still an open and debated issue. We investigate the one-dimensional model, whose microscopical dynamics can be solved exactly, and provide a deeper understanding of the origin of the non-ergodicity. By means of exact arguments, we prove the occurrence of orbits of well-defined periods and their dependence on the conserved energy density. Further statistical estimates of the size of the attraction's basins of the different periodic orbits lead to a complete characterization of the activity vs. energy density phase diagram in the limit of large system's size.
\end{abstract}

\maketitle

Sandpile models have been introduced by Bak, Tang and Wiesenfeld (BTW) as simple non-equilibrium systems exhibiting self-organized criticality (SOC), a concept that provided the theoretical framework for the study of a large number of natural phenomena whose dynamics self-organizes into a critical state in which long-range correlations and scaling laws can be detected~\cite{btw,jensen}. More recently, the occurrence of SOC has been related to the presence of underlying absorbing state phase transitions (APT) governing the critical behavior of the system~\cite{marro}. The relationship between sandpiles and non-equilibrium APTs is emphasized by a class of models with locally conserved dynamics, the Fixed-Energy Sandpiles (FES)~\cite{dickman}. 
Finite size scaling techniques led to the complete characterization of the universality class for driven-dissipative sandpile models with stochastic dynamics~\cite{maya}, but failed in determining the critical properties of the deterministic BTW model, that have been partially recovered by means of multiscaling analysis~\cite{demenech}. 
Stochastic FESs seem to belong to a different universality class with respect to directed percolation~\cite{dickman,vespignani,sven,alava},
while the deterministic BTW dynamics shows very strong non-ergodic effects, that cannot be understood using a purely statistical mechanics approach~\cite{vespignani,bagnoli}. 
In dissipative systems, the non-ergodicity of the BTW rule is related to the existence of recurrent configurations and to the abelian group structure of the configuration space uncovered by Dhar~\cite{dhar}, and emerges when the process of grains addition and dissipation is not completely random. In such cases, the configuration space breaks into separed regions in which the dynamics is governed by cyclic groups and turns out to be periodic~\cite{kurt}. 
In the limit of zero dissipation and addition, the group structure breaks down, but the system becomes similar to a FES, in which the non-ergodic behavior is recovered as the result of a self-sustained activity that eventually enters a periodic orbit. 
 
In this Letter, we report the exact solution of the one-dimensional deterministic Fixed-Energy Sandpile ($1d$-DFES), 
providing an exhaustive description of the microscopic dynamics and its steady states, and determining the system's macroscopic behavior in the limit of large size.  
Moreover, our explanation of the microscopic origin for non-ergodicity in BTW FESs elucidates the relation, already suggested in Ref.\cite{bagnoli}, with mode-locking phenomena in non-linear automata, i.e. when the system's dynamics blocks into orbits of fixed periodicity even though some external parameter (e.g. the energy) is continuously changed in a given interval (see, for instance, Ref.\cite{middleton} and references therein).    

Let us consider a ring of $N$ sites, each one endowed with an {\em energy} $z_{i}$, assuming non negative integer values. Whenever the energy of a site equals or exceeds the threshold value $z^{th}=2$, the site becomes {\em active} and undergoes a toppling event, i.e. it looses $2$ units of energy, that are equally redistributed among its $2$ neighbors. On the contrary, if $z_{i}<2$, the site $i$ is {\em stable}. The system evolves with parallel dynamics, therefore the update rule can be written as follows:
\begin{equation}
z_{i}(t+1) = z_{i}(t) - 2 \theta{\left[z_{i}(t)-2\right]} + \sum_{j \in V_{i}} \theta{\left[z_{j}(t)-2\right]}  
\end{equation}
for $i = 1, 2, \dots, N$, where $\theta(x) = 1 (0)$ if $x \geq 0$ ($x<0$) and $V_{i}$ is the neighborhood of the site $i$.   
When the total energy $E=\sum_{i}z_{i}$ is sufficiently large, the conservation imposed by periodic boundary conditions
prevents the system to relax in a configuration composed of all stable sites ({\em absorbing state}). Since the configuration space is finite and the dynamics is deterministic, after a transient, the system enters a periodic orbit (steady {\em active state}). We will conventionally assume sites to have energy between $0$ and $3$ (relaxing this condition does not modify the physical properties of the system), initial configurations being any possible sequence of $N$ symbols randomly chosen in the alphabet $\{0, 1, 2, 3\}$.
In addition to the invariance under the action of the finite group of cyclic permutations and reflections on the ring, the system admits another internal symmetry: the configurations are dynamically equivalent under the transformation $z_{i} \rightarrow {z'}_{i} = 3 - z_{i}$. This means that the whole orbit traced starting from a configuration $\mathcal{Z}'=\{{z'}_{i}\}$ is known if we know that traced by the configuration  $\mathcal{Z}=\{{z}_{i}\}$, and the two limit cycles have the same period.

In the mean-field description~\cite{dickman,vespignani}, the critical behavior of the system around the absorbing-active transition point is deduced studying the {\em activity phase diagram}, the plot of the {\em activity} $\rho=\frac{1}{N}\sum_{i}\theta(z_{i}-2)$ (the density of active sites) vs. the {\em energy density} $\zeta = E/N$.
The symmetry of the problem allows to restrict the analysis of the $1d$-DFES to the energy density range $\zeta \in [0, 3/2]$ (or in $\zeta \in [3/2, 3]$). 

\begin{figure}[t]
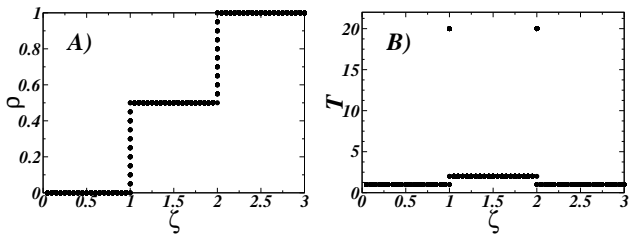

\centerline{
\includegraphics[width=0.225\textwidth]{fig1aFES}~
\includegraphics[width=0.225\textwidth]{fig1bFES}
}
\caption{The activity diagram $\rho (\zeta)$ (A) and the behavior of the period $T$ as function of the energy density $\zeta$ (B) for a $1d$-DFES of size $N=20$. Figures display all results, without averaging, obtained starting from $100$ randomly chosen initial configurations at each energy value.}
\label{fig1}
\end{figure}

Numerical results show that large systems converge to periodic orbits of well-defined periods $T$ in the whole range of the energy density. Fig.~\ref{fig1} reports the diagrams $\rho(\zeta)$ and $T(\zeta)$ for a system of size $N=20$ and $100$ random initial conditions. To emphasize the non-ergodicity, the results of all runs are displayed, without computing any average. In particular, for  $\zeta <1$ the system converges to a completely frozen configuration (period $T=1$) with no active sites, while in the range $1 < \zeta <2$ only orbits of period $T=2$ are observed. As imposed by symmetries, $\zeta >2$ corresponds to fixed-point configurations with all sites active.
At the critical points $\zeta=1, 2$, orbits of period $N$ seem to 
be statistically dominant for large systems, but other periods (e.g. $T=1, 2$) are observed with lower frequency.   

The reasons for such a particular phase-diagram can be understood exploiting a result obtained in the mathematical literature, in which BTW sandpiles have been studied on generic graphs with the name of parallel chip firing games (CFG)~\cite{bjorner}. When the underlying graph is undirected, with neither sinks nor sources, the CFG conserves the total energy. 
To our knowledge, the only result determining the properties of the periodic orbits of CFGs has been proved by Bitar and Goles~\cite{bitar} in the case of trees. Their theorem states that {\em on a tree the steady states of the BTW dynamics are fixed points or cycles of period two}. If the graph contains some loops, the theorem does not hold.
   
Notwithstanding, the method used in Ref.\cite{bitar} can be exploited to study the $1d$-DFES. 
Let us consider a system in a periodic steady state at a time $t_{0}$, and a temporal window ${P}_{t_{0}}=[t_{0}, t_{0}+T-1]$ that corresponds to the first period from $t_{0}$. If the system entered the periodic orbit $\mathcal{O}$ at time $t_{in}$, the whole temporal support $[t_{in},\infty)$ of $\mathcal{O}$ will be indicated with $\mathtt{supp}(\mathcal{O})$. 
We define $s_{i}(t)$ as a binary variable that assumes value $1$ ($0$) if the site $i$ is active (stable) at the time $t$.
The sequence of all these binary values forms the {\em activity vector} $S_{i}= \{s_{i}(t)| t \in \mathtt{supp}(\mathcal{O})\}$ of a site $i$. Moreover, a set $[t,t+p-1] \subseteq \mathtt{supp}(\mathcal{O})$ is a {\em maximal active set} of length $p \geq 1$ if $s_{i}(t+r)=1$ for $r=0,1, \dots, p$ and $s(t-1)=s(t+p+1)=0$.  Similarly, a set $[t,t+q-1] \subseteq \mathtt{supp}(\mathcal{O})$ is a {\em maximal stable set} of length $q\geq 1$ if $s(t-1)=s(t+q+1)=1$ and $s_{i}(t+r)=0$ for $r=0,1, \dots, q$.  

According to Lemma $1$ of Ref.\cite{bitar}, the following statement holds on a generic graph: {\em if $[s-k,s] \subseteq \mathtt{supp}(\mathcal{O})$ is a maximal active (stable) set for a site $i$, then there exists a neighbor $j$ of $i$ such that $[s-k-1,s-1] \subseteq \mathtt{supp}(\mathcal{O})$ is a maximal active (stable) set for $j$}.  
We call $V$ ($W$) the maximum number of consecutive $1$'s ($0$'s) in all the activity vectors along the period, i.e. the length $p$ ($q$) of the largest maximal active (stable) set over all sites. 
The analysis can be restricted to $0<V<T$ ($0<W<T$), the cases $V=0$ ($W=T$) and $V=T$ ($W=0$) corresponding to fixed point configurations in which all sites are active (stable)~\cite{bitar}. 
Moreover, because of the internal symmetry with respect to $\zeta=3/2$, $W$'s properties in the interval $\zeta \in [0, 3/2]$ are equivalent to those of $V$ in the interval $\zeta \in [3/2, 3]$. As a consequence, we limit our study to $W$ and corresponding maximal stable sets in the energy range $\zeta \in [0,3/2]$.  

Two cases, $W=1$ and $W>1$, should be distinguished.    
If $W=1$, the activity vectors of all sites in the system are $2$-periodic.
Suppose, indeed, that a site $i$ topples at a time $t$ together with $m < 2$ of its neighbors: at the following time step $t+1$ the site $i$ does not topple, but the remaining $2-m$ neighbors topple. Thus, having lost $2-m$ energy units during the update at time $t$ and gained the same quantity at the following time step, the value of site $i$ has period $2$. This argument holds for all sites in the periodic state ($W=1$ for all sites), that consequently has global period $T=2$.  

In order to study the case $W>1$, suppose the largest maximal stable set $[t,t+W-1]$ is at a site $i_{0}$. For the above Lemma there is a neighbor $i_{1}$ of $i_{0}$ whose largest maximal stable set is $[t-1,t+W-2]$. In turn, site $i_{1}$ has a neighbor with largest maximal stable set $[t-2,t+W-3]$. However, $W>1$ imposes that $i_{2}\neq i_{0}$, because the Lemma implies $s_{i_{2}}(t-1)=0$ while from the definition of maximal stable set we need $s_{i_{0}}(t-1)=1$.
Proceeding step by step along the ring in the same direction, we reach the site $i_{N}=i_{0}$, whose maximal stable set is $[t-N,t+W-N-1]$. The properties of the system in $i_{0}$ (and in every other site) at time $t-N$ result to be the same as that at time $t$, thus we can conclude that during this process, the system performs one or more periodic cycles and returns in the starting configuration after exactly $N$ temporal steps. In other words, {\em for $W>1$ ($V>1$) the period $T$ divides $N$}.      

\begin{figure}[t]
\begin{center}
\includegraphics[angle=-90,width=0.4\textwidth]{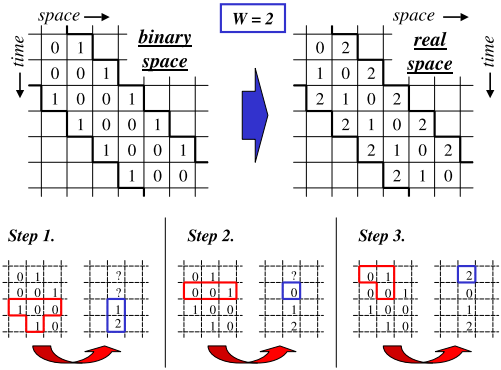}
\end{center}
\caption{Graphical sketch of the method used to find out the structure of periodic configurations. The case $W=2$ is considered. Top figure shows the correspondence between the binary representation of the spatio-temporal dynamics and the  evolution of the real configuration. An example of the procedure used to recover the real values assumed by a given site during the period is reported in the bottom series of draws. Site values can be univocally determined in $W+1$ steps. At each step, the value of the site at a certain time is computed from a number of constraints on its (spatio-temporal) neighbors.}
\label{fig2}
\end{figure}

The rest of this Letter provides a deeper insight into the structure of the periodic orbits using a simple method that is sketched in Fig.~\ref{fig2} for $W=2$.
Suppose that the interval $[t,t+W-1]$ of length $0 < W < N$ is the largest maximal stable set for a site $i$, then $s_{i}(t+s)=0$ for $s=0,\dots, W-1$, but also $s_{i}(t-1)=1$ and $s_{i}(t+W)=1$. In addition, $s_{i-1}(t-1+r)=0$ for $r=0,\dots,W-1$, with $s_{i-1}(t-2)=1$ and $s_{i-1}(t+W-1)=1$; and similarly $s_{i+1}(t+1+r)=0$ for $r=0,\dots,W-1$, with $s_{i+1}(t)=1$ and $s_{i+1}(t+W+1)=1$. These binary values can be drawn on a spatio-temporal grid as in Fig.~\ref{fig2}.
In order to determine the real values assumed by $i$ in the interval $[t,t+W-1]$, we need to discute separately the cases $W=1$, $W=2$ and $W>2$. 

The case $W=1$ is simpler because all maximal stable sets have length $1$, thus $s_{i}(t)=0$ implies that at least one of the two neighbors is active. Drawing a spatio-temporal grid as in Fig.~\ref{fig2}, it follows that spatial configurations can be composed of all possible combinations of two-site blocks of the type $20$, $21$, $30$ and $31$ (and those with opposite order $02$, $12$, $03$ and $13$). 
Note that the set of all possible combinations of these blocks completely fills the energy interval $\zeta \in [1,2]$ of $2$-periodic orbits; while at the energy density $\zeta=1$ ($\zeta=2$), the configurations of the unique $2$-periodic orbit are $\mathcal{Z}_{20}=\{\dots202020\dots\}$ and $\mathcal{Z}_{02}=\{\dots020202\dots\}$ ($\mathcal{Z}_{31}=\{\dots313131\dots\}$ and $\mathcal{Z}_{13}=\{\dots131313\dots\}$). 

When $W>1$ (see Fig.~\ref{fig2}), the site $i$ does not topple at time $t+W-1$ but it has to topple at time $t+W$, after having received a single energy unit (from the site $i-1$) at time $t+W-1$, then $z_{i}(t+W-1)=1$ and, consequently,  $z_{i}(t+W)=2$. 
In particular, if $W=2$, the sites $i$ and $i-1$ do not topple at time $t+W-2=t$, but $i+1$ topples, providing $i$ of the unique unit of energy that we find at time $t+W-1=t+1$. Hence, $z_{i}(t)=0$, from which follows that $z_{i}(t-1)=2$. 

In the case $W>2$, the same arguments show that $z_{i}(t+W-2)=1$ (none of the neighbors provides any grain).  
If we go backward along the maximal complementary set, a site $i$ assumes value $1$ up to a time $t+1$, then $z_{i}(t)=0$ (one of its neighbors topples at that time). 
In summary, during the time interval $[t-1,t+W]$ with $W>1$, the site under study assumes a well-determined sequence of values $2011\dots12$ with exactly $W-1$ values equal to $1$. 

Fig.~\ref{fig2} shows the existing relation between the values assumed by a site during these temporal intervals  and those admitted in the spatial arrangement of the corresponding periodic configurations.    
If the maximal stable set of length $W$ for $i$ starts at time $t$, the uniqueness of the previous construction allows to completely determine the spatial block $\{z_{i-1}(t+W-1) z_{i}(t+W-1) \dots z_{i+W}(t+W-1)\}$ that will be of the form $\{211\dots102\}$ with $W-1$ sites at $1$. Applying a spatio-temporal translation to our construction, the spatial block $\{z_{i-W}(t)\dots z_{i}(t) z_{i+1}(t)\} = \{211\dots 102\}$ is determined as well. 

The remaining structure of the configuration at time $t$ can be studied with a similar technique, provided that we erase the spatial block $\{z_{i-W}(t)\dots z_{i}(t)\}$ and consider a reduced system composed of $N-W-1$ sites. 
Since the role of the erased block was actually only that of transporting an activity `soliton' from the site $i-W$ to the site $i+1$, this operation does not alter the dynamics that maintains a periodic behavior. After the reduction, the above methods are applied on the system, with in principle a largest maximal stable set of different length $W' \leq W$.
At the end of this process, the system (at a time $t$ in the periodic regime) will be decomposed in blocks of decreasing length (from $W+1$ to $2$) and structure of the type $211\dots0$. The only allowed blocks of length $2$ are those of the form $20$.
The fronts direction of motion is determined by the initial conditions, but the dynamics is invariant under spatial reflections. Hence, the analysis can be applied to a system that evolves in the opposite spatial direction to that considered here, producing periodic configurations of the same structure but inverse spatial order. Moreover, the analysis of maximal active sets for $0<V<N$ leads to identical results (in the range $3/2 \leq \zeta \leq 3$) with spatial blocks of the type $122\dots31$ (and $13\dots221$) instead of $211\dots02$ (and $20\dots112$). 

We have proved by construction that the case $W>1$ ($V>1$) corresponds exactly to systems with an energy density $\zeta=1$ ($\zeta=2$). For all other values of $\zeta$, only periods of length $1$ and $2$ are allowed. In the region $\zeta<1$ (and $\zeta>2$) the limit cycles are fixed point configurations with all stable (active) sites, but in the interval $1 < \zeta < 2$ fixed points are forbidden and $W=1$. Then, in this region the period of the limit cycles is $T=2$. Such exact result provides a theoretical foundation to the empirical data collected by numerical simulations (see Fig.~\ref{fig1}). 
\begin{figure}[t]
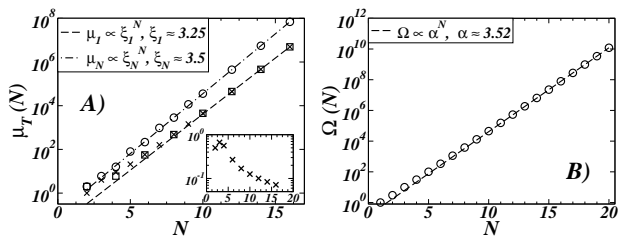

\centerline{
\includegraphics[width=0.22\textwidth]{fig3aFES}~
\includegraphics[width=0.22\textwidth]{fig3bFES}
}
\caption{In panel A), the number $\mu_{T}(N)$ of configurations belonging to the attraction's basin of orbits of period $T=1$ (crosses), $2$ (squares), and $N$ (circles) is plotted as function of $N$. The exponential rate of growth is extimated by curve fitting (dashed line for $\mu_{1}(N)$ and dot-dashed line for $\mu_{N}(N)$). The inset shows that the ratio $\mu_{1}(N)/\mu_{N}(N)$ decreases with $N$.
Panel B) reports the total number $\Omega (N)$ of configurations at $\zeta=1$ (or $\zeta=2$) as function of the system size $N$ (circles). The fit (dashed line) shows an exponential growth as $\Omega(N) \propto \alpha^N$ with $\alpha \simeq 3.5$.}
\label{fig3}
\end{figure}

At the critical points $\zeta=1$ ($\zeta=2$), the knowledge of the dynamics of building blocks allows also to compute the exact number of periodic configuration of period $T$ as a combinatorial enumeration problem.
In particular, the number ${\Pi}_{all}(N)$ of configurations belonging to orbits with $T >1$ equals the number of ways a numbered ring of $N$ sites can be filled with an ordered set of blocks of length $k$ comprised between $2$ and $N$~\cite{flajolet}.
The generating function for the combinations with blocks of length $2\leq k \leq N$ is $Q(z)=1/{\left(1-\sum_{2 \leq k \leq N} z^{k}\right)}$. Hence, ${\Pi}_{all}(N)$ is obtained as function of the coefficients of the derivatives in $z=0$ of $Q(z)$, i.e. ${\Pi}_{all}(N) = \sum_{2 \leq k \leq N}  \frac{2k}{(N-k)!}\frac{d^{(N-k)}}{d {z}^{(N-k)}} Q(z=0) - {A}_{N}$, where $A_{N}=0$ if $N$ is odd and $A_{N} = 2$ if $N$ is even. The last term is introduced to compensate the double counting (due to the factor $2k$) of configurations composed of only $20$ blocks. 
The number ${\Pi}_{T}(N)$ of configurations of period $T<N$ is given by the number of different ways of filling the ring with identical blocks of length and period equal to $T$, i.e. ${\Pi}_{T}(N) = {\Pi}_{T}(T)$ for $2 \leq T <N$.
Formally, ${\Pi}_{N}(N) = {\Pi}_{all}(N) -\sum_{2\leq T < N} {\Pi}_{T}(N)$ (with $T$ divisor of $N$). 
Starting from ${\Pi}_{2}(2)={\Pi}_{all}(2)=2$ and ${\Pi}_{3}(3)={\Pi}_{all}(3)=6$, a recursive relation allows to compute all the other terms ${\Pi}_{T}(N)$ ($2 \leq T \leq N$).  
The explicit computation shows that, in the large $N$ limit, the mass of orbits of period $T=N$ grows faster than that of orbits of periods $T<N$ .

However, in order to establish which period occurs with higher frequency at $\zeta=1$ ($\zeta =2$), the mass of the whole attraction's basin of an orbit is necessary, not only the number of periodic configurations.
We have measured numerically the exact number of configurations in the basins of attraction of orbits of different periods. Fig.~\ref{fig3}-A shows results up to $N=16$. The size $\mu_{1}(N)$ of the basin of attraction of the fixed-point grows exponentially with a rate $\mu_{1}(N)/\mu_{1}(N-1) \simeq \xi_{1}$ with $\xi_{1} \simeq 3.25$. A similar growing rate is observed for orbits of period $T=2$ ($N$ even), while the other orbits of period $T<N$ present smaller attraction's basins. The largest growing rate is that of orbits of period $N$, for which $\mu_{N}(N)/\mu_{N}(N-1) \simeq \xi_{N}$ with $\xi_{N} \simeq 3.5$. This means that for $N=100$ the probability of observing $T=1$ or $T=2$ is about $10^{-4}$ smaller than that of observing $T=N$. 
The conjecture that $N$-periodic orbits have probability $1$ in the large $N$ limit is corroborated by the observation of very similar scaling laws for $\mu_{N}(N)$ and the total number $\Omega(N)$ of possible configurations at energy $E=N$, whose behavior is displayed in Fig.~\ref{fig3}-B (data are computed analytically using simple combinatorics similar to that used in Ref.~\cite{casarta}). 
Consequently, at the critical energies $\zeta=1,2$, with probability $1$ the system enters very long orbits characterized by a steady current of active solitons transported along the system. The activity can assume all (rational) values between $0$ and $1/2$. 

In conclusion, traditional statistical mechanics cannot capture the complex behavior of deterministic FESs, even in one-dimensional systems. On the contrary, our solution provides the full understanding of the mode-locking phenomena and the sharp activity transitions observed in the phase diagram. 
The arguments can be partially extended to higher dimensions, but the analysis is messed up by the complexity of the solitonic motion. In the case of two dimensional lattices, the present analysis together with symmetry arguments allow to obtain an approximated phase diagram~\cite{vivo}, whose structure reproduces the devil's staircase observed in numerical simulations~\cite{bagnoli}.
We hope that these results could also represent a kind of benchmark for the analytical and numerical study of other automata with conserved dynamics, in which spatio-temporal patterns are governed by the motion of solitonic fronts.
  The author is grateful to M.~Casartelli and P.~Vivo for useful comments.

\end{document}